\documentclass[twocolumn,showpacs,preprintnumbers,amsmath,amssymb]{revtex4}
\usepackage{tipa}
\usepackage{graphicx}
\usepackage{dcolumn}
\usepackage{bm}

\begin{document}

\title{Charmonium possibility of $X(3872)$}
\author{Yan-Mei Kong\footnote{kongyanmei@graduate.shu.edu.cn}
and Ailin Zhang\footnote{Corresponding author:
zhangal@staff.shu.edu.cn}}
\affiliation{Department of Physics,
Shanghai University, Shanghai, 200444, China}


\begin{abstract}
Properties of Regge trajectories for charmonium are studied.
Possible interpretations and their implications to newly observed
$X(3872)$ are examined. It is found that the mass of $X(3872)$ is
consistent with the $1^{++}~2^3P_1$ and the $2^{-+}~1^1D_2$
charmonium states.
\end{abstract}
\maketitle

\section{Introduction}

Recently some new charmonium or charmonium-like states, such as
$X(3872)$~\cite{belle,cdf,d0,babar1,cdf2}, $Y(3940)$~\cite{belle1},
$X(3940)$~\cite{belle2}, $Y(4260)$~\cite{babar2,cleo2} and
$Z(3930)$~\cite{belle3} were observed. $Z(3930)$ was pinned down as
the $\chi_{c2}(2p)$ in 2006 PDG~\cite{pdg}, while others have not
been identified. Among these new states, $X(3872)$ has drawn
people's great interest for its peculiar decay properties. $X(3872)$
was first observed by Belle~\cite{belle} in exclusive B decays,
\begin{eqnarray}
B^{\pm} \rightarrow K^\pm X(3872),
X(3872)\rightarrow\pi^{+}\pi^{-}J/\psi.
\end{eqnarray}
Subsequently it was confirmed by CDF~\cite{cdf,cdf2}, D0~\cite{d0}
and BaBar~\cite{babar1}. The mass of this state is $M =
3871.2\pm0.5$ MeV and the width $\Gamma<2.3$ MeV(0.9 C.L.). The mass
is within errors at the $D^0\bar D^{\star 0}$ threshold, but the
width is small.

To accommodate $X(3872)$ in hadron spectroscopy, considerable
speculations and plenty of interpretations were proposed. There are
conventional $c\bar c$ charmonium
assignments~\cite{close,wang,charmonium1,charmonium2,charmonium3,charmonium4,suzuki,charmonium5,charmonium6,godfrey},
molecule state interpretations~\cite{close,mole1,mole2,mole3,mole4},
tetraquark state interpretations~\cite{tetra1}, hybrid
interpretations~\cite{close,hybrid1} or mixing states
interpretations among them~\cite{mole3,mixing1,mixing2}.

Features of hadron were mainly exhibited through its production and
decay properties. Four decay modes of $X(3872)$ have already been
observed. $X(3872)\rightarrow J/\psi\rho$~\cite{belle} and
$X(3872)\rightarrow J/\psi\omega$~\cite{belle4,k} were observed by
Belle, $X(3872)\rightarrow\gamma J/\psi$ was observed by
Belle~\cite{k} and BaBar~\cite{babar3}. Recently, $B\rightarrow
D^0\bar{D^0}\pi^0K$ was observed by Belle~\cite{belle6}.

Possible $J^{PC}$ of $X(3872)$ have been suggested through these
decay modes. The observation of $X(3872)\rightarrow\gamma J/\psi$
indicates that its $C=+$. Analysis of angular distribution in
$X(3872)\rightarrow J/\psi\rho$ favors its
$J^{PC}=1^{++}$~\cite{belle5}. This assignment is also supported by
the observation $B\rightarrow D^0\bar{D^0}\pi^0K$~\cite{belle6}.
However, some analyses suggest that both $1^{++}$ and $2^{-+}$ are
possible~\cite{cdf2,cdf3}.

As for the nature of $X(3872)$, none of the speculations is favored.
In the discovery mode observed by Belle~\cite{belle}, $X(3872)$ was
naturally expected to be the $1^{++}~2^3P_1$ or the $2^{-+}~1^1D_2$
charmonium state due to its decay final states. However, there are
difficulties in the charmonium explanations. $X(3872)$ seems not
match any predicted charmonium state for its lower mass, narrower
width and puzzling decay properties. The upper limit for the
radiative transition $X(3872)\rightarrow\gamma\chi_{c1}$ set by
Belle~\cite{belle} makes it difficult to identify $X(3872)$ with any
charmonium state. The simultaneous decay of $X(3872)$ to
$J/\psi\rho$~\cite{belle} and $J/\psi\omega$~\cite{belle4,k} with
roughly equal branching ratios is a strong implication of the
"molecule" state assignment for $X(3872)$~\cite{mole3,mole4}.

If $X(3872)$ is a "molecule" or tetraquark state, there are also
difficulties. The observed branching fraction of
$X(3872)\rightarrow\gamma J/\psi$~\cite{k,babar3} is much larger
than theoretically predicted one for molecule states~\cite{mole3}.
In particular, if the near-threshold enhancement in $B\rightarrow
D^0\bar{D^0}\pi^0K$~\cite{belle6} is due to $X(3872)$, this mode has
a branching ratio $9.4^{+3.6}_{-4.3}$ times larger than
$B(B^+\rightarrow X(3872)K^+)\times B(X(3872)\rightarrow J/\psi
\pi^+\pi^-)$. This mode appears to be dominant and the branching
ratio is much larger than the predicted one in the molecule model.

So far, there is no compelling evidence to confirm one
interpretation or to exclude one interpretation, more decay modes
have to be searched and studied. Recently, the radiative
$X\rightarrow D\bar D\gamma$ decay mode was
suggested~\cite{vol,colangelo}.

In the charmonium explanation of $X(3872)$, one difficulty is its
lower mass. In quark models, $1^{++}~2^3P_1$ was expected to have
mass $\sim 3953$ MeV in Ref.~\cite{gi} and $\sim 3929$ MeV in
Ref.~\cite{efg}. In general, features of charmonium are described
well by quark models (where the quark dynamics was
assumed)~\cite{heavy}. In lattice, the spectrum of charmonium was
computed also~\cite{latt}. However, the exact quark dynamics in
hadron is not very clear, and lattice results may be improved.
Whether $X(3872)$ really has the lower mass difficulty requires more
phenomenological examinations.

Recently, charmonium spectrum were analyzed in terms of Regge
trajectory theory~\cite{gershtein,ailin1}. However, the analyses
were not complete for limited experimental information at that time.
With more experimental information in hand (the $J^{PC}$ of
$X(3872)$ is believed to be $1^{++}$ or $2^{-+}$ at present time),
we can continue a further exploration of the charmonium possibility
of $X(3872)$ and its implication through its mass relations with
other charmonium states in this Letter.

\section{Regge trajectory and hyperfine splitting of charmonium}

Reggge trajectory~\cite{regge,chew} is an important phenomenological
way to describe the masses relations among different hadrons. There
is resurgent interest in Regge theory for much more accumulated
experimental data. Furthermore, some quark models need more complete
experimental fits for testing~\cite{nonlinear1}. Regge trajectories
are some graphs of the total quantum numbers $J$ versus mass squared
$M^2$ over a set of particles which have fixed principle quantum
number $n$, isospin $I$, dimensionality of the symmetry group $D$
and flavors. A Chew-Frautschi Regge trajectory is a line:
\begin{eqnarray}
J(M^2)=\alpha(0)+\alpha^\prime M^2,
\end{eqnarray}
where intercept $\alpha(0)$ and slope $\alpha^\prime$ depend weakly
on the flavor content of the states lying on corresponding
trajectory. For light quark mesons, $ \alpha^\prime\approx
0.9~GeV^{-2}$. Different Regge trajectories are approximately
parallel.

It is found that the linearity and parallelism of Regge trajectories
with neighborhood mesons stepped by $1$ in $J$ with opposite $PC$
holds not well~\cite{nonlinear2,nonlinear3,ailin2}. The intrinsic
quark-gluon dynamics may result in large non-linearity and
non-parallelism of such Regge trajectories. However, The linearity
and parallelism of Regge trajectories with neighborhood mesons
stepped by $2$ in $J$ with the same $PC$ is found to hold
well~\cite{ani}.

In addition to these properties of the Regge trajectories with the
same principle quantum number $n$, another relation for Regge
trajectories with different $n$ was assumed. It was argued that the
parallelism of Regge trajectories with different $n$ (others are
identical) may hold because of the similar dynamics in
hadron~\cite{nonlinear1}. Whether this parallelism of Regge
trajectory holds has not been tested for the lack of data.

For radial excited light $q\bar q$ mesons, there exist relations
between their masses and principle quantum numbers $n$. These mesons
consist of another kind of trajectory on $(n, M^2)$-plots~\cite{ani}
\begin{eqnarray}\label{nm}
M^2 = M^2_0 + (n-1)\mu^2,
\end{eqnarray}
where $\mu^2$ is the slope parameter(approximately the same for all
trajectories).

Hyperfine (spin-triplet and spin-singlet) splitting relation is
another important mass relation among hadrons. In many potential
models~\cite{potential,potential1,potential2}, the S-wave hyperfine
(spin-triplet and spin-singlet) splitting, $\Delta M_{hf}(nS) =
M(n^3S_1) - M(n^1S_0)$, is predicted to be finite, while other
hyperfine splitting of P-wave or higher L-state is expected to be
zero:
\begin{eqnarray}
\Delta M_{hf}(1P)=<M(1^3P_J)>-M(1^1P_1)\approx 0,\\\nonumber \Delta
M_{hf}(1D)=<M(1^3D_J)>-M(1^1D_2)\approx 0,
\end{eqnarray}
where the deviation from zero is no more than a few MeV. Though
these predictions are model dependent, the masses relation of the
$1P$ charmonium multiplet has been proved to hold in a high degree
accuracy~\cite{pdg}. These hyperfine splitting relations of the
$1P$, $1D$ and $2D$ multiplets will be used as facts (or
assumptions).

These mass relations will be studied or used to explore the
charmonium spectrum. The paper is organized as follows. In the third
section, in terms of the experimental data, all the properties of
possible Regge trajectories for the charmonium are studied, and an
updated phenomenological analysis is made to the new states.
Subsequently, the linearity and parallelism of Regge trajectories
with neighborhood mesons stepped by $2$ in $J$ is combined with the
hyperfine splitting relations of D-wave multiplets to examine some
possible charmonium arrangements to $X(3872)$. Then we analyze
$X(3872)$ through the observed trajectory property on
$(n,M^2)$-plots. Some conclusions and discussions are given in the
last section.

\section{$c\bar c$ possibility of $X(3872)$}

In constituent quark model, $q\bar q$ mesons could be marked by
their quantum numbers, $n^{2S+1}L_J$, and the quantum numbers $PC$
of quarkonia are determined by $P = (-1)^{L+1}$ and $C =
(-1)^{L+S}$. With the most new data for charmonium
mesons~\cite{pdg}, we get table~\ref{table-1}. In this table, the
observed states are listed in the first volume, experimentally
confirmed or favorable theoretical assignment of $J^{PC}$,
$n^{2S+1}L_J$ and masses to these states are put in the sequential
three volumes. Entries in the last volume are information from PDG,
and the states marked with a "?" are those not confirmed or omitted
from the summary table.
\begin{table}
\begin{tabular}{lllllll}
 States & $J^{PC}$ &  $n^{2S+1}L_J$ & Mass(MeV)
& Note\\
\hline\hline $\eta_c(1S)$ & $0^{-+}$ & $1^1S_0$ & 2980.4
& PDG\\
$\eta_c(2S)$ & $0^{-+}$ & $2^1S_0$ & $3638\pm 4$& QN are predictions \\
$\eta_c(3S)$ & $0^{-+}$ & $3^1S_0$ & ? & ? \\
\hline\hline $J/\psi(1S)$& $1^{--}$ & $1^3S_1$ & 3096.9 & PDG \\
$\psi$(2S)& $1^{--}$ & $2^3S_1$ & 3686.1 & PDG \\
$\psi(4040)$& $1^{--}$ & $3^3S_1$ & $4039\pm 1$ & PDG \\
$\psi(4415)$& $1^{--}$ & $4^3S_1$ & $4421\pm 4$ & PDG \\
\hline\hline $\chi_{c0}(1P)$& $0^{++}$ & $1^3P_0$ & 3414.8 & PDG\\
$\chi_{c0}(2P)$& $0^{++}$ & $2^3P_0$ & ? & ? \\
\hline\hline $\chi_{c1}(1P)$& $1^{++}$ & $1^3P_1$ & 3510.7 & PDG\\
$\chi_{c1}(2P)$& $1^{++}$ & $2^3P_1$ & ? & ? \\
\hline\hline $h_c(1P)$& $1^{+-}$ & $1^1P_1$ & 3525.9 & PDG (?, $J^{PC}=?^{??}$)\\
$h_c(2P)$& $1^{+-}$ & $2^1P_1$ & ? & ?\\
\hline\hline$\chi_{c2}(1P)$& $2^{++}$ & $1^3P_2$ & 3556.2 & PDG \\
$\chi_{c2}(2P)$ & $2^{++}$ & $2^3P_2$ & $3929\pm 5\pm 2$ & PDG \\
\hline\hline
$\psi(3770)$& $1^{--}$ & $1^3D_1$ & 3771.1 & PDG \\
$\psi(4160)$ & $1^{--}$ & $2^3D_1$ & $4153\pm 3$ & PDG \\
? & $1^{--}$ & $3^3D_1$ & ? & ? \\
\hline\hline
 ? & $2^{--}$ & $1^3D_2$ & ? & ? \\
 ? & $2^{--}$ & $2^3D_2$ & ? & ? \\
\hline\hline
? & $2^{-+}$ & $1^1D_2$ & ? & ? \\
? & $2^{-+}$ & $2^1D_2$ & ? & ? \\
\hline\hline
? & $3^{--}$ & $1^3D_3$ & ? & ? \\
? & $3^{--}$ & $2^3D_3$ & ? & ? \\
\hline\hline
$X(3872)$ & $?^{?+}$ & ? & 3871.2 & PDG \\
$Y(3940)$ & $?^{??}$ & ? & $3943\pm 11\pm 13$ & PDG (?) \\
$Y(4260)$ & $1^{--}$ & ? & $4259\pm 8^{+2}_{-6}$ & PDG (?) \\
\hline\hline
\end{tabular}
\caption{Spectrum of charmonium.} \label{table-1}
\end{table}

With this table in hand, we can construct different possible Regge
trajectories, study their properties, and proceed with our analysis
of $X(3872)$.

As we know, confirmed states in each group below construct a
trajectory
\begin{eqnarray*}
0^{-+}(1^1S_0),~~~1^{+-}(1^1P_1)\\
0^{++}(1^3P_0),~~~1^{--}(1^3D_1),
\end{eqnarray*}
respectively.

This two trajectories is shown in Fig.1.
\begin
{figure}
\includegraphics{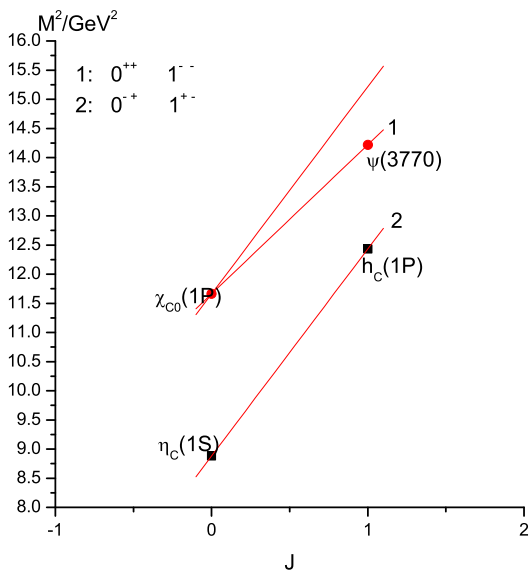}
\caption{Existed trajectories of charmonium singlet and triplet with
$n=1$.}
\end {figure}
In the figure, the slope of line $1$ is $2.558~GeV^2$, the slope of
line $2$ is $3.552~GeV^2$. It's obvious that the two trajectories
are not parallel. Once the parallelism of this two trajectories is
assumed, a large deviation (e.g., $\psi(3770)$ has $0.130~ GeV$
deviation from the "ideal" $1^{+-}(1^1P_1)$ state) would appear.

Another two trajectories with different $n$ are constructed by
$J/\psi(1S)$, $\chi_{c2}(1P)$ and radial excited $\psi(2S)$,
$\chi_{c2}(2P)$, respectively,
\begin{eqnarray*}
 1^{--}~1^3S_1,~~~2^{++}~1^3P_2\\
 1^{--}~2^3S_1,~~~2^{++}~2^3P_2.
\end{eqnarray*}

This two trajectories is shown in Fig.2.
\begin {figure}
\includegraphics{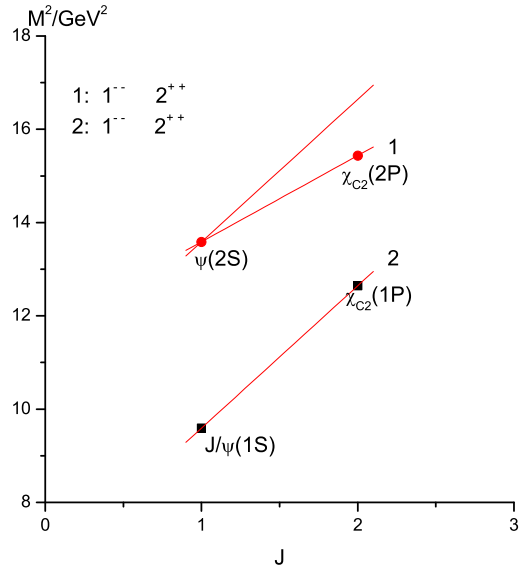}
\caption{Existed trajectories of Charmonium triplets with $n=1$ and
$n=2$.}
\end {figure}
In Fig.2, the slope of line $1$ is $1.850~GeV^2$, the slope of line
$2$ is $3.054~GeV^2$. The discrimination of this two slopes is
obvious, two trajectories are not parallel. The deviation of
$\chi_{c2}(2P)$ from the "ideal" $2^{++}~2^3P_2$ is about
$0.150~GeV$. Obviously, the assumption in Ref.~\cite{nonlinear1}
that the parallelism of Regge trajectories with different $n$  may
hold works not well for charmonium. In fact, even though the
dynamics in hadrons with different $n$ is similar, the parallelism
cannot be deduced directly.

In short, trajectories of charmonium in $(M^2,J)$-plots with
neighborhood mesons stepped by $1$ in $J$ are really not parallel,
and they "fan in"~\cite{nonlinear1}. Therefore, the use of the
parallelism of these trajectories to predict new states is not
reliable if the deviations are unknown.

There exist other Regge trajectories with neighborhood charmonium
stepped by $2$ in $J$. According to Ref.~\cite{ani}, the linearity
and parallelism of this kind of Regge trajectories was found to hold
well. This feature of Regge trajectories for charmonium has not been
tested for the lack of experimental data, it could not give more
predictions either if this feature is used separately. However, once
it is combined with the hyperfine splitting relation in a multiplet,
the $2^{-+}$ ($1^1D_2$ or $2^1D_2$) charmonium possibility of
$X(3872)$ could be examined.

Firstly, let us examine the $2^{-+}~ 1^1D_2$ possibility of
$X(3872)$. In theory, states below construct two Regge trajectories
\begin{eqnarray*}
0^{-+}~ 1^1S_0,~~~~~2^{-+}~ 1^1D_2,\\
1^{--}~ 1^3S_1,~~~~~3^{--}~ 1^3D_3.
\end{eqnarray*}

In this two trajectories, the $0^{-+}~1^1S_0$ and the $1^{--}~
1^3S_1$ are confirmed states, while the $2^{-+}~1^1D_2$ and the
$3^{--}~1^3D_3$ have not been fixed on. If $X(3872)$ is the
$2^{-+}~1^1D_2$ state, the mass of the $3^{--}~1^3D_3$ ($M$) can be
derived in terms of the approximate parallelism relation
\begin{eqnarray}
3.871^2-2.980^2 = M^2-3.097^2
\end{eqnarray}
with $M=3.962~GeV$. In the meantime, the mass of the $2^{--}~1^3D_2$
can be obtained due to zero of hyperfine splitting of the $1D$
charmonium multiplet. The mass of the $2^{--}~1^3D_2$ ($M_2$) is
determined by
\begin{eqnarray}
3.871=\frac{3\times3.771+5M_2+7\times3.962}{15}
\end{eqnarray}
with $M_2=3.804~GeV$, where the spin average is implied.

Therefore, the $1D$ multiplet is  pitched down as follows
\begin{eqnarray*}
1^3D_1~~~~~~1^3D_2~~~~~~1^1D_2~~~~~~1^3D_3~~~~~\\
3.771~~~~~~3.804~~~~3.871~~~~3.962~GeV.
\end{eqnarray*}
The mass of $1D$ spin triplets increases with the increase of $J$,
and the whole mass sequence is reasonable. This analysis implies
that the $2^{-+}$ $1^1D_2$ charmonium arrangement of $X(3872)$ is
compatible with the ordinary mass relation in a multiplet.
Furthermore, the analysis indicates that the $1^3D_2$ is located
around $3.804~GeV$ and the $1^3D_3$ is located around $3.962~GeV$.

The $2^{-+}~2^1D_2$ assignments of $X(3872)$ could be analyzed in a
similar way. In this case, two trajectories are consisted of
\begin{eqnarray*}
0^{-+}~2^1S_0,~~~~~2^{-+}~2^1D_2,\\
1^{--}~2^3S_1,~~~~~3^{--}~2^3D_3,
\end{eqnarray*}
respectively. In this two trajectories, the $0^{-+}~2^1S_0$ and the
$1^{--}~2^3S_1$ are confirmed states, while the $2^{-+}~2^1D_2$ and
the $3^{--}~2^3D_3$ have not been fixed on. If $X(3872)$ is the
$2^{-+}~2^1D_2$ state, the mass of $3^{--}~2^3D_3$ ($M$) is
determined by
\begin{eqnarray}
M^2 - 3.686^2 = 3.871^2 - 3.638^2
\end{eqnarray}
with $M=3.916$ GeV. Once the mass of the $3^{--}~2^3D_3$ is known,
the mass of the $2^{--}$ $2^3D_2$ ($M_1$) is obtained due to zero of
hyperfine splitting of the $2D$ charmonium
\begin{eqnarray}
3.871=\frac{3\times4.153 + 5M_1 + 7\times3.916}{15}
\end{eqnarray}
with $M_1=3.639$ GeV.

The $2D$ spectum are therefore determined as follows
\begin{eqnarray*}
2^3D_1,~~~~2^3D_2,~~~~2^1D_2,~~~~2^3D_3\\
4.153,~~~~3.639,~~~3.871,~~~~3.916.
\end{eqnarray*}
Obviously, the spectrum is exotic ($M(2^3D_1)$ $>$ $M(2^3D_3)$).
That's to say, the $2^{-+}$ $2^1D_2$ charmonium arrangement of
$X(3872)$ seems impossible.

Now, let us study the parallelism of charmonium on $(M^2,n)$-plots.
From table.I, states in each group below construct a trajectory on
$(M^2,n)$-plots,
\begin{eqnarray*}
1^3S_1,~~~2^3S_1,~~~3^3S_1,~~~4^3S_1,~~~\\
1^3P_2,~~~2^3P_2,~~~~~~~~~~~~~~~~~~~~~~~\\
1^3D_1,~~~2^3D_1,~~~~~~~~~~~~~~~~~~~~~~
\end{eqnarray*}
respectively.

This three Regge trajectories on $(M^2,n)$-plots is displayed in
Fig.3.
\begin {figure}
\includegraphics{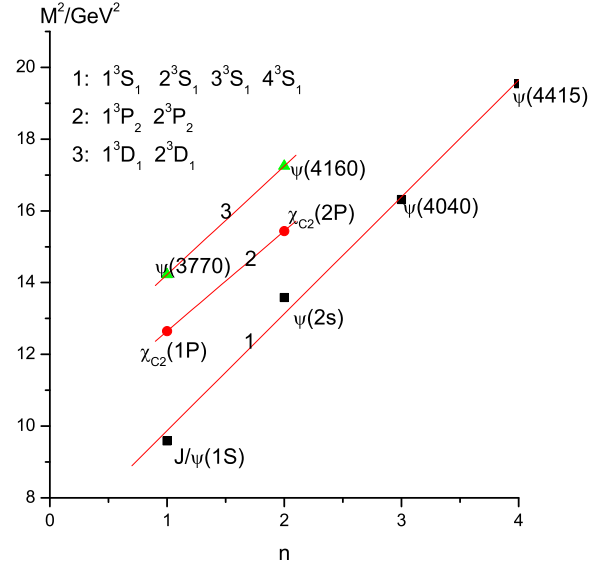}
\caption{Existed trajectories on $(M^2,n)$-plots for $^3S_1$,
$^3P_2$ and $^3D_1$ charmonium.}
\end {figure}
In the figure, the slope of line $1$ is $3.259~GeV^2$, the slope of
line $2$ is $2.792~GeV^2$ and the slope of line $3$ is
$3.027~GeV^2$. It's clear that the difference of the slopes to this
three trajectories is small. These trajectories are approximately
parallel. In terms of this approximate parallelism of Regge
trajectories on $(M^2,n)$-plots, some charmonium assignments to
newly observed states could be examined.

As mentioned in the introduction, $X(3872)$ may be the $1^{++}$
$2^3P_1$ candidate. $Y(3940)$ may be the $2^3P_0$~\cite{gershtein}
or the $3^1S_0$~\cite{charmonium2} candidate. If $X(3872)$ is the
$1^{++}~2^3P_1$ and $Y(3940)$ is the
$0^{++}~2^3P_0$~\cite{gershtein}, states in each group below will
construct a trajectory on $(M^2,n)$-plots:
\begin{eqnarray*}
1^3S_1,~~~2^3S_1,~~~3^3S_1,~~~4^3S_1,~~~\\
1^3P_2,~~~2^3P_2,~~~~~~~~~~~~~~~~~~~~~~~\\
1^3D_1,~~~2^3D_1,~~~~~~~~~~~~~~~~~~~~~~\\
1^3P_1,~~~2^3P_1,~~~~~~~~~~~~~~~~~~~~~~~\\
1^3P_0,~~~2^3P_0.~~~~~~~~~~~~~~~~~~~~~~~
\end{eqnarray*}
This five trajectories is plotted in Fig.4. If the assignments to
these states are correct, five Regge trajectories should be
approximately parallel due to previous arguments.
\begin {figure}
\includegraphics{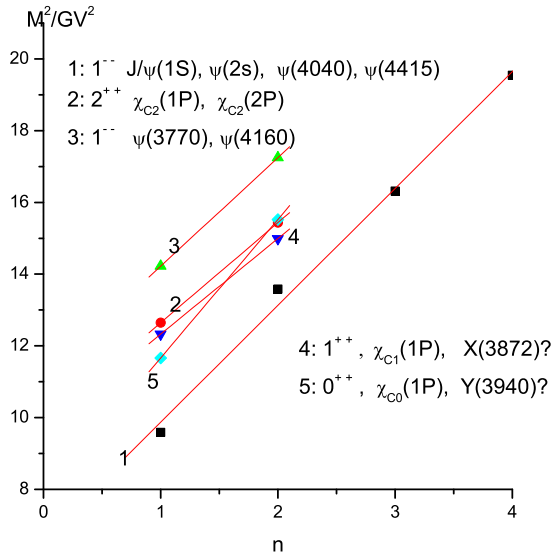}
\caption{Possible trajectories on $(M^2,n)$-plots for the $^3S_1$,
$^3P_2$, $^3D_1$, $^3P_0$ and $^3P_1$ charmonium.}
\end {figure}

The slope of line $1$ is $3.259~GeV^2$, the slope of line $2$ is
$2.792~GeV^2$, the slope of line $3$ is $3.027~GeV^2$, the slope of
line $4$ is $2.665~GeV^2$, and the slope of line $5$ is
$3.861~GeV^2$. In this figure, it is easy to observe that the
trajectory $5$ (with $Y(3940)$ involved) intersects with
trajectories $2$ and $4$, while the trajectory $4$ (with $X(3872)$
involved) approximately parallels trajectories $1$, $2$ and $3$.
From these observations, it is reasonable to conclude that the
$1^{++}$ $2^3P_1$ charmonium suggestion for $X(3872)$ does not
contradict with possible mass relations in charmonium. As a
byproduct, the $2^3P_0$ charmonium assignment of $Y(3940)$ seems
impossible.

$X(3872)$ may be a four-quark state ($[cq][\bar c\bar q]$ tetraquark
state or molecule
state)~\cite{close,mole1,mole2,mole3,mole4,tetra1}, but the
four-quark state possibility of $X(3872)$ will not be studied here.
Four-quark states have been extensively studied for a long
time~\cite{jaffe,chan,isgur1,ted,okun,bander,glashow,isgur2,tetra1,jaffe2,maiani1,tetra1,ebert,
stech}, unfortunately, their properties especially their dynamics
and decay properties are still unfamiliar. So far, many states such
as $f_0(600)$ (or $\sigma$), $f_0(980)$, $a_0(980)$, the unconfirmed
$\kappa(800)$, $D^\ast_{SJ}(2317)^\pm$, $X(3872)$, $Y(4260)$,
$X(1835)$ and $X(1812)$~\cite{pdg,4260,1812} have once been
interpreted as four-quark state, but no one has been confirmed.

\section{Conclusions and discussions}

The nature of $X(3872)$ is still unclear. In addition to its
$J^{PC}$($1^{++}$ or $2^{-+}$), whether it is a charmonium state or
an exotic state is still uncertain. Instead of the production and
decay properties, some mass relations of $X(3872)$ are studied in
the charmonium assignments. Through these relations and some
phenomenological analyses, some assignments of $X(3872)$ are
examined.

If $X(3872)$ is a $2^{-+}$ state, it may be the $2^{-+}~1^1D_2$
charmonium state while is unlikely to be the $2^{-+}~2^1D_2$
charmonium state. If it is really the $2^{-+}~1^1D_2$ charmonium
state, the whole $1D$ multiplet is pitched down with the $1^3D_2$
located around $3.804~GeV$ and the $1^3D_3$ located around
$3.962~GeV$.

If $X(3872)$ is a $1^{++}$ state, it may be the $1^{++}~2^3P_1$
charmonium. As a byproduct, it is found that $Y(3940)$ is unlikely
to be the $0^{++}~2^3P_0$ charmonium.

So far, the study of four-quark state and the study of meson near
thresholds are not satisfactory. Four-quark state is usually invoked
to explain the special decay properties of newly observed states,
which in fact may be explained also without four-quark
state~\cite{bugg}. Only when the properties of four-quark state are
definitely clear, the four-quark state explanation of newly observed
state will be satisfactory.

Of course, some properties in our analyses may not have firm
foundation. The linearity and parallelism of Regge trajectories with
neighborhood mesons stepped by $2$ in $J$ may be questionable (the
deviations may be relevant to the spin-dependent interactions in
quark models), however, the analyses here provide a complementary
study of $X(3872)$. These properties could be tested by more
forthcoming experimental data and may give hints to the quark
dynamics in hadron.

Acknowledgment: This work is supported in part by Shanghai Leading
Academic Discipline Project with project number: T0104.

\end{document}